\documentstyle[preprint,aps]{revtex}
\tighten
\draft
\begin{document}
\title{Novel Edge Excitations of Two-dimensional Electron Liquid in a
Magnetic Field}
\author{I. L. Aleiner, and L. I. Glazman}
\address{Theoretical Physics Institute,
School of Physics and Astronomy, University of Minnesota,\\
116 Church Str. SE, Minneapolis, MN 55455}
\maketitle

\begin{abstract}
We investigate the low-energy spectrum of excitations of a
compressible electron liquid in a strong magnetic field. These
excitations are localized at the periphery of the system. The
analysis of a realistic model of a smooth edge yields new branches of
acoustic excitation spectrum  in  addition to the well known edge
magnetoplasmon mode.   The velocities are found and the observability
conditions are established for the new modes.
\end{abstract}

\narrowtext
The dispersion relation for plasmons in a non-restricted
two-dimensional electron liquid is well known to have a form
$\omega\propto k^{1/2}$ \cite{Ritchie,Stern,Chaplik}. If the liquid
has a boundary, an edge mode appears in addition to these bulk
excitations. The spectra of the edge and bulk modes differ from each
other only by a numerical factor \cite{Volkov}. A magnetic field
applied perpendicularly to the plane of the liquid changes the
plasmon spectrum drastically. The spectrum of the bulk mode acquires
a gap of the width equal to the cyclotron frequency $\omega_c$. The
only known gapless mode existing in the presence of the magnetic
field propagates along the boundary
\cite{Volkov,Mast,Glattli,Fetter}. The ``chirality'' of this edge
magnetoplasmon  determined by the direction of the magnetic field (i.
e., by the sign of the Hall conductivity $\sigma_{xy}$) was
demonstrated explicitly in the time-domain experiments \cite{Ashoori}.

The solved theoretical models of the edge modes assumed a sharp
electron density profile at the
boundary\cite{Volkov,Mast,Glattli,Fetter}, i.~e. width of the
boundary strip was assumed to be infinitesimal. The existence of only
a single branch of the edge magnetoplasmons follows directly from
this assumption. For a realistic shape of a potential confining the
electron liquid, the density profile is smooth at the
boundary\cite{Glazman,Chklovskii,Gelfand}. The results of
Refs.\cite{Volkov,Mast,Glattli,Fetter} can be extended on this case
only under the assumption that the current and charge  oscillations
forming the magnetoplasmon wave are homogeneous across the boundary
strip. However, the latter condition is excessively restrictive. We
demonstrate in this paper the existence of other sound-like modes
propagating along the edge. The current for each of these modes
alternates across the boundary strip, and therefore the new branches
could not be predicted on the basis of a ``sharp'' boundary model.

Below we present an exactly solvable model correctly describing all
the edge excitations in the strong magnetic field limit. We obtain
also the values of the oscillator strengths and the damping of these
modes. The new branches become robust and are not destroyed by a
finite relaxation time in the achievable region of relatively short
wavelengths.

The dynamics of the compressible electron liquid is governed by Euler
equation and the continuity equation linearized in the velocity of
the liquid $\mbox{\boldmath $v$}\left(\mbox{\boldmath
$\rho$},t\right)$ and in the deviation of the concentration $\delta
n\left(\mbox{\boldmath $\rho$},t\right)$ from its equilibrium value
$n_0\left(\mbox{\boldmath $\rho$}\right)$:
\begin{eqnarray}
 &\dot{\mbox{\boldmath $v$}} +
 \omega_c  \left( \hat{\mbox{\boldmath $z$}} \times
 \mbox{\boldmath $v$}\right) -
 \frac{e^2}{\varepsilon m}
\mbox{\boldmath $\nabla_{\rho}$}
 \int d^2\mbox{\boldmath $\rho$}_1
\frac {\delta n \left(\mbox{\boldmath $\rho$}_1\right)}
    {|\mbox{\boldmath $\rho$} -\mbox{\boldmath $\rho$}_1|}
 = 0&, \label{hydro} \\
 &\delta\dot{n}
 +  \mbox{\boldmath $\nabla_{\rho}$}\left(n_0 \mbox{\boldmath
$v$}\right) = 0.&
 \label{motion}
\end{eqnarray}
Here {\boldmath $\rho$} is radius-vector in the plane $XY$ of the
two-dimensional electron liquid, $\hat{\mbox{\boldmath $z$}}$ is the
unit vector along $Z$-axis, and $\varepsilon$ is the dielectric
constant. The last term in (\ref{hydro}) represents Coulomb
interaction\cite{meanfield}.

In the following we assume that the electron liquid is homogeneous in
$y$ direction and occupies half-plane $x>0$. Since the system is
translationally invariant in the $y$ direction, we will seek the
solution of Eqs.~(\ref{hydro}), (\ref{motion})  in the form
\begin{eqnarray}
&\mbox{\boldmath $v$}
 =\exp\left(iky-i\omega t\right)\mbox{\boldmath $w$}(x);&
\nonumber \\
&\delta n(x,y) = \exp\left(iky-i\omega t\right)f(x).&
   \label{seek}
\end{eqnarray}
Substituting Eqs.~(\ref{seek}) into the system (\ref{hydro}),
(\ref{motion}) and eliminating $\mbox{\boldmath $w$}(x)$, we find an
integral equation for $f(x)$:
\begin{equation}
\left(\omega_c^2-\omega^2\right)f
+\frac{2e^2}{\varepsilon m}\left\{k^2 n_0 - n_0\frac{d^2}{dx^2}-
n_0^{\prime}\frac{d}{dx}+
\frac{k}{\omega} \omega_c n_0^{\prime}\right\}
\int_0^{\infty} K_0\left(|k||x-x_1|\right)f(x_1)dx_1= 0,
\label{basic}
\end{equation}
where $n_0^{\prime}\equiv dn_0/dx$ and $K_0(x)$ is the modified Bessel
function. Homogeneous equation (\ref{basic}) comprises the eigenvalue
problem that determines the spectrum of edge  excitations $\omega_j
(k)$. The spectrum is controlled by the parameters of the problem: by
the magnetic field determining $\omega_c$, and by the concentration
profile $n_0(x)$. The latter depends on a particular type of the
confining potential\cite{Glazman,Gelfand}. We are interested in the
low-frequency, long wavelength modes, and this allows us to neglect
the terms proportional to $\omega^2$ and $k^2$ in Eq.~(\ref{basic}).

Futher simplifications are possible in the case of a strong magnetic
field. Keeping only the terms proportional to $\omega_c^2$ and
$\omega_c$,  and introducing a new function
\begin{equation}
g(x)= \left(\frac{dn_0}{dx}\right)^{-1/2}f(x)
\label{g}
\end{equation}
instead of $f(x)$, we find from (\ref{basic}) the following reduced
equation:
\begin{eqnarray}
 g(x)&=&\lambda \int_0^{\infty}
K_0\left(|k||x-x_1|\right)\frac{1}{\bar{n}}
\sqrt{\frac{dn_0}{dx}\frac{dn_0}{dx_1}}g(x_1)dx_1,
\label{Fredholm}\\
\omega &=&-\frac{2}{\lambda} \frac{\bar{n}e^2}{\varepsilon m
\omega_c}k. \label{lambda}
\end{eqnarray}
Here $\bar{n} \equiv n_0(x \to \infty)$ is the density of the
homogeneous electron liquid  far from the boundary. One can estimate
from Eqs.~(\ref{Fredholm}), (\ref{lambda}) the typical value of
$|\omega/k|$ to be of the order of $\bar {n}e^2/m\omega_c$. It
follows also from (\ref{g}), (\ref{Fredholm}) that the charge
distribution $f(x)$ in the wave is localized mainly within the region
where $dn_0/dx$ is large, i.~e. within the boundary strip of width
$a$. Now we can establish the validity criterion of the strong
magnetic field approximation. The neglected in Eq.~(\ref{basic})
terms $\propto \bar {n}/a $ are smaller than the main terms $\propto
m\omega_c^2/e^2$ if the condition
\begin{equation}
\omega_c^2 \gg \frac{\bar {n}e^2}{\varepsilon m a}
\label{condition}
\end{equation}
is satisfied. For realistic parameters of the two dimensional
electron system formed in a GaAs heterostructure \cite{Chklovskii},
$\bar {n}\sim 1/a_B^2$ and $a \sim 10 a_B$, the latter condition is
equivalent to a rather weak restriction on the filling factor, $\nu
\lesssim 10$.

Eq. (\ref{Fredholm}) is the integral equation of Fredholm type. Its
kernel is symmetric and positively defined, hence, all the
eigenvalues $\lambda$ are real and positive. If one makes an
approximation $K_0(|k||x-x_1|)\approx\ln\left(1/|ka|\right)$ leading
to the degeneracy of the kernel, then only a single finite eigenvalue
$\lambda$ exists. This eigenvalue corresponds to the known
magnetoplasmon mode\cite{Volkov}. The actual kernel in
(\ref{Fredholm}) is non-degenerate, however, and thus there are many
edge modes.

The eigenvalue problem (\ref{Fredholm}) can not be solved
analytically for an arbitrary distribution $n_0(x)$. Below we present
a model for the density profile :
\begin{equation}
 n_0(x) = \frac{2}{\pi}\bar{n}\arctan\sqrt{\frac{x}{a}}, \quad x \geq
0,
\label{n}
\end{equation}
that allows a complete analytical solution of the problem. This model
describes correctly the asymptotic behavior of the density formed by
an electrostatic confinement, reproducing the
characteristic\cite{Glazman,Chklovskii,Gelfand}
$\sqrt{x}$-singularity at $x \to 0$.

Proposed model allows us to solve
the eigenvalue problem (\ref{Fredholm}) using
an expansion of $g(x)$ in a form
\begin{equation}
 g(x)=\frac{1}{(x)^{1/4}(x+a)^{1/2}}\sum_{j=0}^{\infty}g_j
      T_{2j} \left(\sqrt{\frac{a}{x+a}}\right),
\label{expansion}
\end{equation}
where  $T_{n}(\xi)\equiv\cos (n\arccos\xi)$ are the Chebyshev
polynomials\cite{Abramovitz}. Substitution of  (\ref{expansion}) into
(\ref{Fredholm}) leads to the following system of equations for
coefficients $g_j$ of the expansion:
\begin{eqnarray}
\frac{1}{\lambda} g_0 &=&\ln \left(\frac{e^{-\gamma}}{2|ka|}\right)
g_0- \sum_{j=1}^{\infty}\frac{(-1)^j}{j}g_j,\nonumber\\
\frac{1}{\lambda} g_j &=&\frac{1}{j}g_j  - 2\frac{(-1)^j}{j}g_0,
\quad j\geq 1.
\label{coefficients}
\end{eqnarray}
When deriving Eqs.~(\ref{coefficients}), we used the approximation
$K_0(kx)\approx\ln\left(2e^{-\gamma}/|kx|\right)$ which is valid in
the long-wavelength limit, $|ka| \ll 1$; here $\gamma\approx
0.577...$ is the Euler constant. The system (\ref{coefficients})
leads directly to the following transcendent equation for the
eigenvalues:
\begin{equation}
\frac{1}{\lambda} + 2\Psi\left(1-\lambda\right)=
\ln\left(\frac{e^{-3\gamma}}{2|ka|}\right).
\label{spectrum}
\end{equation}
Here $\Psi(1-\lambda)$ is the digamma function\cite{Abramovitz} that
has simple poles at $\lambda = 1, 2,\dots$. Because $|ka| \ll 1$, the
solutions of Eq.~(\ref{spectrum}) are close to the points $\lambda=0,
1, 2, \dots$ where the left-hand side of this equation is singular.
The smallest root of Eq.~(\ref{spectrum}) belongs to the region
$\lambda \ll 1$. Expanding the left-hand side of this equation in
power series in $\lambda$ and retaining only the two leading terms of
the expansion, we find the spectrum of the conventional\cite{Volkov}
edge magnetoplasmon mode:
\begin{equation}
\omega_0(k) = -2\ln \left(\frac{e^{-\gamma}}{2|ka|}\right)
\frac{\bar{n}e^2}{\varepsilon m\omega_c}k.
\label{plasmon}
\end{equation}
Other roots $\lambda \geq 1$ are close to the poles of
the digamma function in
Eq.~(\ref{spectrum}). It is these roots that determine the new
branches of the edge excitations with acoustic spectrum:
\begin{equation}
\omega_j(k) =-s_jk,\quad s_j= \frac{2\bar{n}e^2}{\varepsilon m
\omega_cj}, \quad j=1,2,\dots.
\label{acoustic}
\end{equation}

The  difference between the acoustic modes and the ``usual'' plasmon
($j=0$) originates in  the structure of charge distributions
associated with these waves. In the usual plasmon wave, charge does
not oscillate across the boundary strip,  whereas in the acoustic
mode $j$ charge oscillates $j+1$ times in $x$ direction (see Fig.~1),
the average density being smaller by a factor of $\left| j\ln
(|ka|)\right|^{-1}$ than the oscillations amplitude for each mode
with $j\neq 0$. The potential energies $U_j$ produced by the charge
distribution types depicted in Fig.~1 can be easily estimated. For
the same characteristic amplitudes of charge density perturbation in
all the waves, we find the ratio  $U_0/U_j \simeq j|\ln(|ka|)|$. This
explains the difference between the spectra (\ref{plasmon}) and
(\ref{acoustic}), as energies $U_j$ provide the restoring forces for
the modes. For the higher harmonics $j\gg1$ the latter
considerations  are obviously independent on the particular density
profile (\ref{n}), that allows one to expect certain universality of
the spectrum (\ref{acoustic}). Indeed, for an arbitrary profile
$n_0(x)$, the function
\begin{equation}
g(x)=\left(\frac{dn_0}{dx}\right)^{1/2}\cos\left(\pi j \frac{n_0(x)
}{\bar{n}}\right)
\label{neutral}
\end{equation}
may be used at $j \gg 1$ as an asymptotic solution of the eigenvalue
problem (\ref{Fredholm}). In the case of the profile derived in
Ref.~\cite{Glazman}, we find the corrections to the velocities $s_j$
(see Eq.~(\ref{acoustic})) to be of the order of $1/j^2$.

The observability of the  acoustic modes requires sufficiently large
oscillator strengths and rather slow decay for these modes. To begin
with, we evaluate the oscillator strengths $S_j^{\alpha \beta}(k)$
for all the modes; here $\alpha, \beta=x,y$ denote the polarization
of the applied AC electric field,
$E_{\alpha}(y,t)=E_{\alpha}\exp(iky-i\omega t)$, and $j=0,1,2,\dots$
is the mode number. The power $P$ absorbed from the AC field within
the unit length of the boundary is related to the oscillator
strengths of the different modes by:
\begin{eqnarray}
P &\equiv& \frac{e}{2L}\mbox{Re}\int n_0\mbox{\boldmath $vE^*$}
d^2\mbox{\boldmath $\rho$} \nonumber\\
&=& \frac{1}{2}\sum_{j=0}^{\infty} S_j^{\alpha \beta}(k)
E_{\alpha}E_{\beta}^* \delta (\omega - \omega_j(k)).
\label{P}
\end{eqnarray}
Here $L$ is the length of the boundary, and velocity \mbox{\boldmath
$v$} is the linear response to the external electric field
\mbox{\boldmath $E$}. To calculate \mbox{\boldmath $v$}, one has to
add the term $e\mbox{\boldmath $E$}/m$  to the right-hand side of the
equation of motion (\ref{hydro}).  The use of Eqs. (\ref{hydro}),
(\ref{g}) - (\ref{lambda}) and (\ref{P}) allows to express
$S_j^{\alpha \beta}(k)$ in terms of the eigenfunctions $g_j(x)$.  We
present here the results for the diagonal components of the
oscillator strengths:
\begin{equation}
S_j^{\alpha\alpha}(k)=\frac{\pi\omega_j(k)\bar{n}e^2}{m\omega_ck}
F_j^{\alpha}(k),
\label{S}
\end{equation}
\[
F_j^{x}=
\left(k\int dx\frac{n_0}{\bar{n}}
\sqrt{\frac{\bar{n}}{n^{\prime}_0}}g_j\right)^2, \quad
F_j^{y}=\left(\int dx
\sqrt{\frac{n^{\prime}_0}{\bar{n}}}g_j\right)^2,
\]
functions $g_j(x)$ here being normalized by condition
\begin{equation}
\int dx g_j^2(x)=1.
\label{norm}
\end{equation}
It is obvious from Eq. (\ref{g}) that $F_j^{y}$ is proportional to
the square of the average charge mode $j$ bears. As was mentioned
already, this charge in the acoustic modes is parametrically smaller
than the one in the usual edge magnetoplasmon mode. Therefore, the
interaction of AC field  polarized along the boundary with the
acoustic modes is much weaker than the interaction with the
magnetoplasmon.  An explicit calculation in the framework of our
model gives:
\begin{equation}
 S_j^{yy}=\frac{1}{\varepsilon}\left(
\frac{\bar{n}e^2}{m\omega_c}
\right)^2 \times
\left\{
\begin{array}{ll}
2\pi\left|\ln\left(|ka|\right)\right|,&j=0;\\
4\pi\left|\ln\left(|ka|\right)\right|^{-2}j^{-3},&j\geq1.
\end{array}
\right.
\label{S1}
\end{equation}
AC electric field applied perpendicular to the boundary interacts with
the $x$ component of the dipolar moment of the modes. These moments,
and correspondingly factors $F_j^x$  are of the same order of
magnitude for all the modes. Therefore, the difference in absorption
for this polarization is only due to the difference in the mode
frequencies:
\begin{equation}
 S_j^{xx}=\frac{1}{\varepsilon}\left(
\frac{\bar{n}e^2}{m\omega_c}
\right)^2 \times
\left\{
\begin{array}{ll}
2\pi^3\left|\ln\left(|ka|\right)\right|^{-1},&j=0;\\
4\pi^3\left|\ln\left(|ka|\right)\right|^{-2} j^{-1},&j\geq1.
\end{array}
\right.
\label{S2}
\end{equation}
It is interesting to notice that the absorption anisotropies
$S_j^{xx}/S_j^{yy}-1$ are of the opposite signs for the usual plasmon
and for the acoustic modes respectively.

To estimate the decay rates of different edge modes, we include a
phenomenological relaxation time $\tau$ into the equation of motion
(\ref{hydro}) by the substitution $\dot{\mbox{\boldmath $v$}} \to
\dot{\mbox{\boldmath $v$}} + \mbox{\boldmath $v$}/\tau$. After such a
modification the energy of the mode $\epsilon_j$ becomes time
dependent, $\epsilon_j(t)\propto\exp(-2t/\tau_j)$. The latter
relation allows one to define the relaxation rates $1/\tau_j$. For
small dissipation the result can be expressed in terms of the
unperturbed eigenmodes $g_j(x)$ :
\begin{equation}
\frac{1}{\tau_{j}}=\frac{\omega_j(k)}{\omega_c\tau k} \int dx n_0(x)
\left(\frac{d}{dx}\frac{g_j}{\sqrt{n_0^{\prime}}}\right)^2,
\label{tau}
\end{equation}
$g_j(x)$ being normalized by the condition (\ref{norm}). As one can
easily see from Eq.~(\ref{tau}), the relaxation rate increases with
the mode number because of oscillatory behavior of the
eigenfunctions. We find for the plasmon and acoustic modes:
\begin{equation}
\omega_j\tau_j = |ka|\omega_c\tau\times\left\{
\begin{array}{ll}
2\left(\ln |ka| \right)^2, &j=0;\\
\beta_j/j^2,  &j\geq 1,
\end{array}
\right.
\label{hard}
\end{equation}
where $\beta_1=6/5$, $\beta_2=60/53$,\dots, $\beta_{\infty}=1.128$
are slowly varying with $j$ numerical factors.

Solutions of Eq.~(\ref{Fredholm}) and perturbative results
(\ref{hard}) obtained above are applicable as long as dissipation is
small enough so that it does not affect the charge distribution in
the eigenmodes. The characteristic length of the charge spreading
$l_{\omega}$ caused by dissipation is inversely proportional to
frequency, $l_{\omega}=e^2\bar{n}/(\varepsilon
m\omega\omega_c^2\tau)$.  The smallness of the redistribution
requires $l_{\omega}$  to be shorter than the characteristic length
scale $\sim a/(j+1)$ for spatial variations of the eigenfunctions
$g_j(x)$. The latter condition imposes different restrictions for the
plasmon and acoustic modes; with the help of (\ref{plasmon}),
(\ref{acoustic}), we find: $|ka|\ln(1/|ka|)\gtrsim 1/(\omega_c\tau)$
for $j=0$ and  $|ka|\gtrsim j^2/(\omega_c\tau)$ for $j\neq 1$.  At
smaller wavevectors, the results of Volkov and Mikhailov\cite{Volkov}
for the spectrum and decay rate are applicable\cite{matching},
whereas the acoustic modes are over-damped. The region of
observability of the new branches is shown on Fig.~2.

The observability condition for the new modes is quite restrictive.
Indeed, as it follows from (\ref{hard}), the product $\omega_c \tau$
must be at least greater than $1/|ka|$ for the first acoustic mode to
be observed. If the characteristic values of $k$ are determined by a
sample perimeter (which is typically $\sim 1$ cm\cite{review}), the
condition (\ref{hard}) can not be satisfied even for the mobility of
$10^6$ $\mbox{cm}^2/\mbox{V sec}$. The technique using a metallic
grating coupler\cite{Demel} appears to be more promissing. In such an
experiment the wave vector $k=2\pi/d$ is determined by the grating
period $d$ that may be made\cite{Demel} of the order of 1
$\mu\mbox{m}$. For the typical width of the boundary strip $\sim
2000\AA$ this implies the condition $\omega_c\tau\gtrsim 1$.

In conclusion, we found the new low-frequency excitations propagating
along the edge of a two-dimensional electron liquid in the presence
of a magnetic field. These new modes have acoustic spectra with
velocities inversely proportional to the mode indices. At a given
wavevector $k$, the frequencies of acoustic modes are lower than that
of a conventional edge magnetoplasmon by a factor $1/|\ln ka|$, where
$a$ is the width of the boundary strip. The oscillator strengths and
decay times for the acoustic branches with low indices differ from
the corresponding values for the magnetoplasmon by powers of the same
small parameter. The logarithmic function appears due to the
long-range nature of Coulomb interaction. If the electron system is
confined by gate-induced potentials, the differences in the mentioned
parameters for the acoustic modes and the conventional magnetoplasmon
become less significant: the logarithmic function should be replaced
by a factor of the order of unity because of the screening of Coulomb
interaction by the metallic gates.

The authors are grateful to H.~U. Baranger, K. ~A. Matveev, V. ~I.
Perel and B. ~I. Shklovskii for helpful  discussions and comments.
This work was  supported  by NSF Grants DMR-9117341 and DMR-9020587.

\begin{figure}
\caption{Characteristic charge distributions for: (a) the edge
magnetoplasmon mode, $j=0$; and (b) the edge acoustic mode with $j=2$.
Charge patterns shown in the figure move along the $Y$-axis
according Eq.~(\protect\ref{seek}).}
\end{figure}
\begin{figure}
\caption{The first three branches of the edge excitations;
$\omega_*=2\bar{n}e^2/(\varepsilon m\omega_c a)$. The dashed line
separates the regions of strong damping (below the line) and weak
damping (above the line). In the latter region the acoustic edge
modes become observable.}
\end{figure}

\begin{references}
\bibitem{matching} Formulas for the spectrum and damping
($\omega_0\tau_0= (2/\pi)\ln\omega_c\tau$) in the low-frequency
regime of\cite{Volkov} can be obtained by replacing $a$  in
(\ref{plasmon}) and (\ref{hard}) with a length $\sim l_{\omega}$.
\end{references}
\end{document}